\documentstyle[amssymb]{article}

\def\al{\alpha}
\def\ga{\gamma}
\def\om{\omega}
\def\la{\lambda}
\def\iy{\infty}
\def\De{\Delta}
\def\de{\delta}
\def\si{\sigma}
\def\iy{\infty}

\def\dac{\displaystyle\frac}

\def\dil{\displaystyle\int\limits}

\begin{document}

\begin{center}

\vspace{1cm}

{\large{\bf Spectroscopy through the change of undulator parameters in
dipole regime

\vspace{1cm}

M.L. Schinkeev}}

\medskip

\it{Tomsk Polytechnic University}

\end{center}

\vspace{2cm}

{\bf Abstract}

\medskip

In this work a method of spectroscopy without monochromators for the
undulator radiation (UR) source is proposed. This method is based of
changing the magnetic field in the undulator. Different variations of field
modulations and corresponding object reaction for the case of a dipole
regime of UR exitation were considered. The results of a numerical
experiment are shown and possibilites of this method for an undulator
consisting of two blocks and rearranging by changing the distance between
the blocks were estimated.

\medskip

{\bf 1. Introduction}

\medskip

In recent works [1-3] it has been proposed to use undulator radiation for
spectroscopy without monochromators. According to these proposals the UR
without an intermediate converter falls on the spectroscopical object,
where the spectral density of the UR flux is multiplied with the spectral
sensitivity of the detector on every wavelength  $\la$ and then summed.
 This summation gives an object reaction on the UR flux integrally over all
 wavelengths. Considering the UR spectrum can be relativistically changed
 (by changing the particles energy or the undulator field), the reaction
 of the object on the rearranged source of radiation can be defined as an
 integral Fredgolm equation of the $I$ kind:
\begin{equation}
I(P)=\dil_0^\iy {K(P,\la)A(\la)d\la},
\label{lab1}
\end{equation}
where $I(P)$ is the object reaction on the UR flux, $K(P,\la)$ is the
equation kernel (\ref{lab1}) - spectral density of the radiation flux,
$A(\la)$ is the spectral function of the radiacion acceptor, and $P$ is the
rearranging parameter. The determination of the object function $A(\la)$ by
the measured reaction $I(P)$ reduces to the solution of the integral
equation with the known kernel $K(P,\la)$.

As was shown in Ref. [\ref{ref2}], the simplest way to solve this equation
is to use the energy of the accelerated particles as a rearranging
parameter. This method was already used in Refs. [\ref{ref2},\ref{ref3}].
However, the active change of the particle energy is effectively possible
only for accelerators and rather problematic by big storage rings, to which
the majority of users are now oriented. Spectroscopy without monochromators
attract interest, because in this case the spectral properties of the
source could be changed not by the particle energy, but by changing the
undulator field.

\medskip

{\bf 2. Basic relations}

\medskip

Singe spectroscopy without monochromators produces no loss in intensity of
the source on an intermediate converter, it is suitable to use an undulator
with the dipole regime of UR exitation for the dipole parameter $k^2\ll1$.
By this means, spectroscopy without monochromators is confined to changes
in the UR field structure (not taking into consideration changes of
$k^2$), but  it gives, nowever, a simler possibility for the numerical
realization of the algorithm of the solution of Eq. (\ref{lab1}). The
spectral density of the photon flux in an arbirary spectral interval
$d\bar\om=d\om/\om$ for a radiating charge in the dipole approximation,
according to Refs[\ref{ref2},\ref{ref3}], can by written as:
\begin{equation}
\dac{d\Phi}{d\bar\om}=\dac{8\pi\al}{e}
\Big(\dac{\mu_0 e}{\pi mc}\Big)^2 J \eta
\dil_\eta^\iy
{H^2(\nu)\Big(1-2\dac{\eta}{\nu}+2\dac{\eta^2}{\nu^2}\Big)\dac{d\nu}{\nu^2}},
\label{lab2}
\end{equation}
where $J$ is the magnitude of the current of the accelerated charge in the
storage ring, $\al$ is the fine structure constant, $e$ is the electron
charge, $m$ is the electron mass, $\mu_0$ is the magnetic permeability of
vacuum, $\eta=(\la_0/(2\la\ga^2))(1+k^2)$ is the number of the UR harmonic;
$\ga$ is the Lorentz factor, $\la_0$ is the period length of the undulator
magnetic field, and $H^2(\nu)$ is the square modulus of the
Fourier-structure of the UR magnetic field.

As was shown in Ref. [\ref{ref3}] for spectroscopy without monochromators an
undulator consisting of periodical blocks is suitable. Its Fourier
structure of the field can be written as follows:
\begin{equation}
H(\nu)=G(\nu)\Psi(\nu)S(\nu),
\label{lab3}
\end{equation}
where $G(\nu)$ is the Fourier structure of a standart undulator element
(UE) (i.e. an element from which an undulator half-period is formed),
$\Psi(\nu)$ is the Fourier structure of an undulator block, as a set of UE,
$S(\nu)$ is the Fourier structure of the undulator, as a set blocks.
$G(\nu)$ is the expression for an ironless electromagnetic undulator with
the winding of an optimal profile section [\ref{ref3}]:
\begin{equation}
G(\nu)=\dac{j\la_0^2}{2} \dac{e^{-2\pi\nu h/\la_0}}{(\pi\nu)^2}
\Big(\cos\Big(\dac{\pi\nu}{4}\Big)-\cos\Big(\dac{\pi\nu}{2}\Big)-
\dac{\pi\nu}{4}\sin\Big(\dac{\pi\nu}{4}\Big)\Big),
\label{lab4}
\end{equation}
where $j$ is the current density in winding section UE and $2h$ is the
magnetic gap. For $n$-fold balanced charge motion in the undulator block
(this corresponds to the condition of turning to zero of the $n$-fold
integral from the UR field along undulator block), with period number $N$,
the Fourier structure is [\ref{ref3}]:
\begin{equation}
\Psi(\nu)=\bigg(2\cos\Big(\dac{\pi}{2}(1+\nu)\Big)\bigg)^n
\cdot\dac{\sin\Big(\dac{\pi}{2}(1+\nu)(2N-n)\Big)}
{\sin\Big(\dac{\pi}{2}(1+\nu)\Big)}.
\label{lab5}
\end{equation}
If all $M$ undulator blocks are equal and the distances between
neighbouring blocks are $l$, then for $S(\nu)$ one can derive from
Ref. [\ref{ref3}]:
\begin{equation}
S(\nu)=2\cos\Big(\dac{\pi}{2}(\de+\nu(2N+L))\Big)
\cdot\dac{\sin\Big(\dac{\pi}{2}(\de+\nu(2N+L))M\Big)}
{\sin\Big(\dac{\pi}{2}(\de+\nu(2N+L))\Big)},
\label{lab6}
\end{equation}
where $L=2l/\la_0$, i.e. it defines the distances between the blocks in
units of half-periods, and parameter $\de$ defines the phasing of the
blocks switching on: the value $\de=0$ correspondes to the in phase
switching on; the value $\de=1$ correspondes to switching oninthe opposite
phase.

As was shown in Ref. [\ref{ref3}] for the purpose of monochromator-free
spectroscopy the integral spectrum, taken as the difference in the integral
spectra corresponding to the undulator state with in phase and opposite
phase switching on, is most suitable as the kernel of Eq. (\ref{lab1})
(further, the differential UR kernel). Since for such a kernel the low
frequency range of the spectrum is suppressed and accordingly the
contributions of "off-axis" particles into the radiation are also
suppressed, the $\sigma$-component polarization appears. Further, by
$S(\nu)$ the expression corresponding to the phase state difference is
meant.

\medskip

{\bf 3. Object reaction on UR flux}

\medskip

Undulators for obtaining arbitrary UR are described in detail in the
literature. It is possible to modulate the properties of the UR flyx for a
multiblock undulator, as described above, either by changing the
geometrical parameters of the UE, or by changing the contribution to
undulator of the separate elements, or by changing the blocks location.

If the field is modulated by changing the geometrical parameters of the UE,
then the most technically realized, in this case, is changing of the
magnetic gap. However, if we consider the dipole exitation UR regime, then
the corresponding changing interval $h$ does not essentially change the
form of the spectrum. This is clearly seen if one considers the integral UR
spectrum according to Eq. (\ref{lab2}) in the first UR harmonic
approximation. In this case (\ref{lab2}) becomes
\begin{equation}
\dac{d\Phi}{d\bar\om}=CG^2(h)\eta
\dil_\eta^\iy{\Psi^2(\nu)S^2(\nu)
\Big(1-2\dac{\eta}{\nu}+2\dac{\eta^2}{\nu^2}\Big)\dac{d\nu}{\nu^2}}
\end{equation}
($C=$const.) and integra loperator $K(h,\eta(\la))=d\Phi/d\bar\om$ is
degenerate because it can by written as $K(h,\eta)=K_1(h)K_2(\eta)$. If we
consider a set of harmonics, for example the first and the third, then
changing of $h$ leads to a rearrangement of the contributions of these
harmonics in the general spectrum.

Changing the contributions of the UE in the frame of expression
(\ref{lab5}), is really possible only for an electromagnetic undulator
system, and even this is possible only in discrete limits. Therefore in
this work this case is not considered.

So, the most simple variant of UR  field modulation is modulation by
changing the undulator blocks location, i.e., by changing the distance $L$
between the blocks. It is obvious that an arbitrary $L$ can by realized for
$L>0$ (by mechanical removement), because for $L<0$ only discrete values
exist (by rearranging the contributions into the undulator).

The reaction of an object with spectral sensitivity $A(\la)$ by the full
capture of the UR flux over all wavelengths changable over the undulator
$L$ is defined as follows:
\begin{equation}
I(L)=\dil_\eta^\iy{A(\la)\dac{d\Phi}{d\bar\om}(\eta(\la),L)\dac{d\la}{\la}}
=\dil_\eta^\iy{A(\eta)\dac{d\Phi}{d\bar\om}(\eta,L)\dac{d\eta}{\eta}}
=\dil_\eta^\iy{A(\eta)K(\eta,L)d\eta}.
\label{lab7}
\end{equation}
At $L\rightarrow\iy, I(L)$ goes asymptioticaly to 0; that is why changing
the interval of $L$ is defined by giving the reaction level $I(L)$. On the
other side, real limitations on changing of $L$ exist. They are correlated
with the value of the intervening space in the insertion device. Since the
limits of changing of $L$ define the method resolution
($\De\la/\la\approx1/L_{max}$),at a certain value of the intervening space
$L_{max}$, it is better to use blocks with minimal values $N$, which
guarantee a meaningful level $I(L)$ over the whole interval, and for making
the radiation characteristics better one has to use the maximal balancing
degree of the block $n=N-1$. A solution of Eq. (\ref{lab7}) $A(\eta)$ for
the measured reaction $I(L)$ and the analytically given equation kernel
$K(\eta,L)$ can be found by means of the Tichonov regularization method
[\ref{ref4}].

\medskip

{\bf 4. Numerical model of the algorithm}

\medskip

Since the resolution $\De\la/\la$ of the metod of defining the object
spectral function by its reaction (\ref{lab7}) is a value of the order of
$1/L_{max}$ (where $L_{max} \sim N$) and the number of calculation
operations grows proportionally to prodict $NL_{max}$, then for a simple
realization of a numerical model an undulator consisting of two blocks was
used; there were four elements in every block with a balancing degree of
every block $n=3$. The corresponding contribution structure of the
exitation currents of the blocks elements is written as [\ref{ref3}]
$\{-1;3;-3;1\}$.

As a function of spectral sensitivity of object $A(\eta)$ a set of Gaussian
was used:
\begin{equation}
A(\eta)=\sum_{i=1}^2{e^{(\eta-\eta_i)^2/{\si_i}^2}},
\end{equation}
with parameters $\eta_i=\{0.6;0.9\}$, $\si_i=\{0.3;0.1\}$, so that the
reflection $\eta=\la_0/(2\la\ga^2)$ gives the function:
\begin{equation}
A(\la)=\sum_{i=1}^2{e^{-(1-\la_i/\la)^2/{\bar\si_i}^2}},
~~\bar\si_i=\si_i/\eta_i,
\end{equation}
corresponding to two spectral lines with the relation $\De\la/\la$ with is
equal to 0.35 and 0.08 respectively. (Relatively small values of
$\De\la/\la$ correspond to a choice of the undulator parameters made
above).

The kernel of the equation was defined over the interval $\eta\in(0,2)$,
and the limits of changing of $L$ were taken from 0 to 20. The size of the
net, which approximates the kernel of Eq. (\ref{lab7}) was taken equal
to $(200\times400)$.

To solve Eq. (\ref{lab7}) the Tichonov regularization method with a choice
of the regularization parameter according to a generalization of
nonbuilding principle [\ref{ref4}] was used. For the minimization of
Tichonov functional the method of gradients was used.

\medskip

{\bf 5. Conclusion}

\medskip

From the results of the numerical experiments one can see that the
considered algotithm of monochromator-free spectroscopy may by rather
perspective for insertion devices where an active changing of the particles
energy is problematic. For a realization of this method a simple undulator,
consisting of two blocks and rather small period number $N$ in every block
$(\approx 5 - 20)$ but with the opportunity to change the distances between
the blocks, is necessary. The theoritical resolution of this method is
inversely proportional to the distance between the blocks $L_{max}$ (in
units of half-periods). The object reaction
practically disappears at a shift value $L\approx10N$.  This value,
probably, defines the maximally possible method resolution by a given period
number $N$ in the block. So, if we consider a period number for each block
$N=20$ and a period length of 2 cm, then we will obtain $L_{max}\approx200$
and accordingly the possible resolution,
$\De\la/\la\approx1/L_{max}=5\cdot10^{-3}$ (by the total length of
such a twoblock system), is equal $(2N+L_{max})\cdot2=480$ cm.

\medskip

{\bf References}

\begin{enumerate}
\item M.M. Nikitin and G. Zimmerer.
{\it Use of undulators for spectroscopy without monochromators.}
 -- Nucl. Instr. and Meth. A 240 (1985) 188.\label{ref1}
\item V.G. Bagrov, A.F. Medvedev, M.M. Nikitin and M.L. Shinkeev.
{\it Computer analysis for undulator radiation spectroscopy without
monochromators.}
 -- Nucl. Instr. and Meth.  A 243 (1987) 156.\label{ref2}
\item V.F. Zalmeg, A.F. Medvedev, M.L. Shinkeev and V.Ya. Epp.
{\it On the construction of block periodic undulators with metrologically
pure radiation kernels}
 -- Nucl. Instr. and Meth. A 308 (1991) 337.\label{ref3}
\item A.N. Tichonov et al.
{\it Regularization Algorithms and Information.}
 -- Nauka, Moskow, 1983, in Russian.\label{ref4}
\item M.M.Nikitin and V.Ya.Epp. {\it Undulator Radiation.}
 -- Energoatomizdat, Mos\-kow, 1988, in Russian.\label{ref5}
\end{enumerate}
\end{document}